\documentclass[prd,twocolumn,showpacs,letterpaper,nofootinbib]{revtex4}
\usepackage{amsmath}
\usepackage{graphicx}
\usepackage{graphics} 
\usepackage{color} 

\usepackage{amssymb}

\newcommand{\be}{\begin{equation}}
\newcommand{\ee}{\end{equation}}
\newcommand{\bea}{\begin{eqnarray}}
\newcommand{\eea}{\end{eqnarray}}

\begin{document}

\title{Gibbons-Hawking Boundary Terms and Junction Conditions for Higher-Order
Brane Gravity Models}
\author{Adam Balcerzak}
\email{abalcerz@wmf.univ.szczecin.pl}
\author{Mariusz P. D\c{a}browski}
\email{mpdabfz@wmf.univ.szczecin.pl}
\affiliation{Institute of Physics, University of Szczecin,
Wielkopolska 15, 70-451 Szczecin, Poland.}

\date{\today}

\begin{abstract}

We derive the most general junction conditions for the fourth-order brane gravity
constructed of arbitrary functions of curvature invariants. We reduce these
fourth-order theories to second order theories at the expense of introducing new
scalar and tensor fields - the scalaron and the tensoron. In order to obtain junction
conditions we apply the method of generalized Gibbons-Hawking boundary terms which are appended to the appropriate actions. After assuming the continuity of the scalaron and the tensoron on the brane, we recover junction conditions for such general brane universe models
previously obtained by different methods. The derived junction conditions can serve 
studying the cosmological implications of the higher-order
brane gravity models.

\end{abstract}


\maketitle

\section{Introduction}
\label{sect0}
\setcounter{equation}{0}

In the simplest approach to the variational principle of any
theory of gravity, such as Einstein relativity, one used to assume
that both the variation of the metric tensor $\delta g_{ab}$, and
the variation of the first derivative of the variation of the metric tensor $\delta g_{ab;c}$,
vanish on the boundary of the integration volume \cite{wald}.
However, motivated by quantum cosmology, Gibbons and Hawking \cite{GH}
claimed that the latter requirement is too strong, and so they suggested exact
cancellation of the term coming from the variation of the action
which involves the derivatives with a postulated
extra boundary term -- nowadays called just the Gibbons-Hawking boundary term.
For Einstein relativity the
Gibbons-Hawking boundary term is constructed of the trace of the
extrinsic curvature \cite{GH}. It is expected that any theory of gravity
may be appended by an appropriate boundary term. Hawking and
Lutrell \cite{lutrell} found a boundary term for the fourth order
gravity theory composed of the combination of the square of the
Weyl tensor and an arbitrary function of the scalar curvature in four dimensions and
studied its Wheeler-deWitt quantization. The boundary terms
for gravity theory of an arbitrary function of the scalar curvature were also
studied by Barrow and Madsen \cite{madsen} and
in a more general case of an arbitrary function of curvature invariants
by Barvinsky and Solodukhin \cite{barvinsky}.
The study of the appropriate boundary terms
for the Gauss-Bonnet density being one of the general Lovelock
densities \cite{lovelock,briggs} has also been done \cite{bunch81,surface,davis,gravanis}.
These densities are, however, special combinations of the curvature invariants which give the second-order field equations. The Gibbons-Hawking boundary term for Lovelock gravity with AdS asymptotics has been found, too \cite{olea}.

Brane gravity theories initiated by Ho\v{r}ava and Witten
\cite{hw} and further developed by Randall and Sundrum \cite{RS}
add an extra aspect to the problem of boundary terms, since the
boundary of the bulk spacetime is the brane, which serves as the
actual universe \cite{brane}. In analogy to a surface layer
problem of electromagnetism, in brane models, one has to study
appropriate junction conditions \cite{visser}. The problem of the
gravitating surface layer junction conditions was first solved by
Israel \cite{israel66}, and recently applied to brane universes in
Gauss-Codazzi formalism \cite{GaussCod}. It was also solved for
a Brans-Dicke braneworld \cite{abdalla}. All these considerations became
the basic conditions for studying cosmological solutions within the framework of brane
scenario which allows the modification of the Newton's law on small scales.
Brane models produce a different cosmological framework, since the square in energy density
term and the dark radiation resulting from transferring the gravitons from
the bulk to the brane and vice versa appear in generalized cosmological equations \cite{GaussCod}.

Within the physically interesting context, it seems challenging to formulate
braneworld scenario for generalized gravity theories such as Lovelock gravity
and the fourth-order gravity. In order to achieve that, one necessarily has to
formulate Israel junction conditions for these gravities on the brane.
In fact, for the Gauss-Bonnet brane, these conditions were formulated by the
application of the Gauss-Codazzi formalism in Ref.
\cite{deruelle00}. This allowed many detailed studies of the
Gauss-Bonnet brane cosmologies \cite{charmousis,jim,lidsey,maeda},
and of general Lovelock cosmologies \cite{meissner01}. However,
Lovelock brane models are free from the problem of divergencies
resulted from the appearance of the powers of the delta function
in the field equations \cite{meissner01,paper1}. In fact, gravity theories which are
based on the lagrangians being the functions
of curvature invariants such as $f(R)$ theory (see \cite{f(R)};
for a recent review see \cite{RMP}) or $f(R,R_{ab}R^{ab},R_{abcd}R^{abcd})$
theory \cite{clifton,quadratic}, unavoidably lead to such divergencies and
the formulation of the junction conditions is a non-trivial task. In Ref.
\cite{paper1} we have proposed the resolution of the problem
for such theories by imposing more regularity
onto the metric tensor at the brane position, though still keeping the theory to be
a fourth-order. However, such a strong regularity of the metric at the brane may
seem somewhat restrictive and so we have also explored the equivalence of
these conditions to the conditions obtained for an equivalent
second-order theory with an extra scalar degree of freedom - the
scalaron \cite{afrolov}. We have also assumed the continuity of the scalaron
on the brane. Such a correspondence for surface layer
$f(R)$ universes has already been studied in Refs.
\cite{borzeszkowski,branef(R)} and quite recently in Ref.
\cite{deruelle07} in the Gauss-Codazzi approach. On the other hand,
in Ref. \cite{braneR2} the gravity theory of the linear combination
$f(R,R_{ab},R_{abcd})=aR^2 + bR_{ab}R^{ab} +cR_{abcd}R^{abcd}$
($a,b,c=$ const.) was studied in the Gibbons-Hawking boundary term
approach. In these references some example cosmological solutions have been found.
Our current task is to extend these considerations onto $f(R,R_{ab}R^{ab},R_{abcd}R^{abcd})$ theory by using the Gibbons-Hawking boundary term method, also in the most general case, where the discontinuity of the new fields -- the scalaron and the tensoron -- at the brane position
is allowed.

In Section \ref{sect1} we discuss junction conditions for the fourth-order $f(R)$ gravity
theory (after transforming it to the second order
theory) by adding an appropriate Gibbons-Hawking boundary term, constructed of the extrinsic
curvature and an extra scalar field - the scalaron. In Section \ref{sect2} we derive junction conditions for a general $f(R,R_{ab}R^{ab},R_{abcd}R^{abcd})$ gravity theory, first 
by transforming it to the second-order theory, and then by adding an appropriate Gibbons-Hawking boundary term, which is constructed of the extrinsic curvature and an extra rank four tensor field - the tensoron. In fact, we benefit from the discussion of general junction conditions for these higher-order theories following our complementary
approach given in Ref. \cite{paper1}, but we derive more general junction conditions --
the ones which do not possess continuity of the scalaron and the tensoron at the brane. In Section \ref{sum} we give our conclusions.

\section{Gibbons-Hawking boundary term and junction conditions for f(R) gravity}
\label{sect1}
\setcounter{equation}{0}

As the first example of the application of the Gibbons-Hawking boundary term method
to derive junction conditions for brane universes, we discuss the $f(R)$ gravity theory
in $D$ spacetime dimensions
\bea
\label{ac}
S_{p} &=& \chi^{-1} \int_{M_p} d^D x \sqrt{-g} f(R)+ S_{bulk,p}~,
\eea
where $R$ is the Ricci scalar, $\chi$ is a constant, $S_{bulk,p}$ is the bulk matter
action, and $p=1,2$. It is known \cite{branef(R)} that this theory gives
fourth-order field equations and that it can be expressed in an equivalent way by using
the following action
\begin{eqnarray}
\label{equiv2}
\bar{S}_{p} &=& \int_{M_p} d^D x \sqrt{-g}\{f'(Q)(R-Q) + f(Q)\} \\ \nonumber &+&
S_{bulk,p} ~,
\end{eqnarray}
where $M_{p}$ is the spacetime volume, $Q$ is an extra field (a lagrange multiplier), $K$
is the trace of the extrinsic curvature, $h$ is the determinant of the induced metric and $f'(Q) = df(Q)/dQ$. Varying (\ref{equiv2}) with respect to $Q$, one obtains an equation of motion $Q=R$ (provided $f''(Q) \neq 0$), and this is equivalent to a scalar-tensor Brans-Dicke gravity with Brans-Dicke parameter $\omega_{BD}=0$, and the potential $V(H) = -HQ(H) +
f(Q(H))$ (see e.g. \cite{afrolov}). Defining $H=f'(Q)$ (a scalaron
\cite{afrolov}) one can rewrite (\ref{equiv2}) in the form
\begin{eqnarray}
\label{equiv3}
\bar{S}_{p} &=& \int_{M_p} d^D x \sqrt{-g}\{HR - V(H)\} + S_{bulk,p}~.
\end{eqnarray}
The variation of (\ref{equiv3}) gives
\begin{eqnarray}
\label{var}
 \delta \bar{S}_{p} &=&\int_{M_p} d^D x
\sqrt{-g} \left\{ \left[ {1\over 2} g^{ab} HR - HR^{ab} \right.\right. \\
\nonumber &-& \left. \left.
{1\over 2}g^{ab}V(H)+ H_{;dc}(g^{ab}g^{cd}-g^{(ac}g^{b)d}) \right.\right. \\ \nonumber
&-& \left.\left. {\chi \over 2} T^{ab}_{~~bulk,p} \right]\delta g_{ab} \right.
+ \left. \left(R+ {\partial V \over \partial H}\right)\delta H \right\} \\ \nonumber
&+&\int_{\partial M_p}d^{D-1}x \sqrt{-h}2H \underline{n_{d}}_{;c}g^{a[b}g^{c]d} \delta g_{ab} \\ \nonumber
&+&\int_{\partial M_p}d^{D-1}x \sqrt{-h}4H_{;(c}\underline{n_{d)}}g^{a[b}g^{c]d}\delta g_{ab}
\\ \nonumber
&-& \int_{\partial M_p}d^{D-1}x\sqrt{-h}(2Hg^{a[b}g^{c]d} \underline{n_{d}} \delta
g_{ab})_{;c}~,
\end{eqnarray}
where $\underline{n^{a}}$ is a unit vector normal to $\partial M$, and it is chosen to be
``outward pointing'', if spacelike,
and ``inward pointing'', if timelike (this is according to the Gauss integral theorem),
and the induced metric is given by $h_{ab}=g_{ab}-\epsilon n_{a}n_{b}$
($\epsilon=1$ for a timelike brane, and $\epsilon=-1$ for a spacelike brane).
In our case the brane divides spacetime $M$ into two parts: $M_1$ and $M_2$. This
makes convenient the usage of the symbol $M_p$ to indicate integral domains
of (\ref{var}) (this is why $p=1$ or $p=2$). It should be noticed that
$\underline{n^{d}}$ changes its direction into the opposite, if the calculations are done
on the other side of the brane (that is if $p$ changes its value from 1 to 2), while
the vector $n^{a}$ remains the same.

Let us now briefly discuss this problem in a more detailed way.
At first, assume that $n^{a}$ is ``outward pointing'' for $p=1$. Such an assumption
implies that $n^{a}$ is an ``inward pointing'' for $p=2$. If,
for $p=1$, the vector $n^{a}$ is spacelike (the case of a timelike brane),
then $\underline{n^{a}}$ is an ``outward pointing'', and so
$n^{a}=\underline{n^{a}}$, $n^{a}\underline{n_{a}}=1$.
If, for $p=1$, the vector $n^{a}$ is timelike (the case of a spacelike brane),
then $\underline{n^{a}}$ is ``inward pointing'', and so
$n^{a}=-\underline{n^{a}}$, $n^{a}\underline{n_{a}}=1$.
Similarly, if for $p=2$, the vector $n^{a}$ is spacelike
(the case of a timelike brane), then $\underline{n^{a}}$ is ``outward pointing'', and so
$n^{a}=-\underline{n^{a}}$ (because in the case of $p=2$, $n^{a}$ is ``inward pointing''),
$n^{a}\underline{n_{a}}=-1$. If, for $p=2$, the vector $n^{a}$ is timelike
(the case of a spacelike brane), then $\underline{n^{a}}$ is ``inward pointing'', and so
$n^{a}=\underline{n^{a}}$, $n^{a}\underline{n_{a}}=-1$. It should also be stressed
that, for $p=1$, we have $h^{ac}\underline{n_{d}}_{;c}=\epsilon K^{a}_{d}$, and
for $p=2$, we have $h^{ac}\underline{n_{d}}_{;c}=-\epsilon K^{a}_{d}$ \cite{wald}.
Unifying these considerations one can write that
\begin{eqnarray}
n^{a}&=&-(-1)^{p}\epsilon \underline{n^{a}}~, \nonumber
\\ \nonumber
n^{a}\underline{n_{a}}&=&-(-1)^{p}~,  \\ \nonumber
h^{ac}\underline{n_{d}}_{;c}&=&-(-1)^{p}\epsilon
K^{a}_{d}~.
\end{eqnarray}\\
The last term of the formula (\ref{var}) contains the first derivatives of
the metric variation $\delta g_{ab;c}$, and in the Gibbons-Hawking approach \cite{GH} these derivatives should not necessarily be imposed to vanish. Instead, one adds
an appropriate boundary term, which cancels the derivatives. The problem is to find its explicit form. In order to do so, let us first choose
the foliation fulfilling the condition that $n^{a}n_{b;a}=0$. Using this,
one can derive the following identity
\begin{eqnarray}
\label{di}
X^{c}_{~;c}= D_{e}\{h^{ea}X_a\} + \epsilon K n_{a}X^{a} +
\epsilon \mathcal{L}_{\vec{n}}(n_{a}X^{a})~,
\end{eqnarray}
where  $D_{e}$ is a covariant derivative on the brane and
${\mathcal{L}_{\vec{n}}}$ is the Lie derivative in the direction of the vector field $\vec{n}$.
If we take $X_c$ as
\begin{equation}
X^{c}=2Hg^{a[b}g^{c]d} \underline{n_{d}} \delta g_{ab}~,
\end{equation}
then we have
\begin{eqnarray}
\label{nX}
n_{a}X^{a}= -(-1)^{p}H h^{ab}\delta g_{ab}~.
\end{eqnarray}
With the help of (\ref{nX}), the formula (\ref{di}) can be expressed as
\begin{eqnarray}
\label{di2}
X^{c}_{~;c}&=& D_{e}\{h^{ea}X_a\} -(-1)^{p} \epsilon K H h^{ab}\delta g_{ab} \nonumber
\\ \nonumber
&-&
(-1)^{p} \epsilon \{\mathcal{L}_{\vec{n}}H\}h^{ab}\delta g_{ab} \\ \nonumber
&-&
(-1)^{p} \epsilon H \{\mathcal{L}_{\vec{n}}h^{ab}\}\delta g_{ab}  \\
&-&
(-1)^{p} \epsilon H h^{ab}\{\mathcal{L}_{\vec{n}}\delta g_{ab}\}~.
\end{eqnarray}
Using the identities
\begin{eqnarray}
\mathcal{L}_{\vec{n}}h^{ab}&=&-2 K^{ab}, \\ \nonumber
h^{ab}\{\mathcal{L}_{\vec{n}} g_{ab}\}&=& 2 K,  \\ \nonumber
h^{ab}\{\mathcal{L}_{\vec{n}}\delta g_{ab}\}&=& h^{ab}\delta
\{\mathcal{L}_{\vec{n}} g_{ab}\} +
\epsilon K n^{a}n^{b} \delta g_{ab} \\ \nonumber &-& 2D_{b}\{h^{b}_{a} \delta
n^{a}\},
\\ \nonumber
\delta h^{ab}\{\mathcal{L}_{\vec{n}}g_{ab}\}&=&-2K^{ab}\delta g_{ab}, \\ \nonumber
\delta \sqrt{-h}&=& {1 \over 2} \sqrt{-h} h^{ab} \delta g_{ab},
\end{eqnarray}
and inserting them into (\ref{di2}), after integrating out pure divergencies, we obtain:
\begin{eqnarray}
\label{di21}
X^{c}_{~;c}&=& -(-1)^{p}\sqrt{-h}\{ \epsilon h^{ab} H_{;c}n^{c}
\nonumber \\ \nonumber &+& H K n^{a} n^{b}
+ 2\epsilon n^{b} h^{ae}D_{e}H \} \delta g_{ab}\nonumber \\ &+& 2(-1)^{p}\sqrt{-h}
\epsilon K \delta H \nonumber \\  &-&
\delta \{ 2 (-1)^{p} \sqrt{-h}\epsilon H K\}~,
\end{eqnarray}
The last term in (\ref{di21}) is exactly what we need to cancel
the derivatives of the metric variation $\delta g_{ab;c}$. In fact, this is
exactly the Gibbons-Hawking boundary term which should be appended to the action (\ref{ac})
\cite{lutrell}. It reads as
\begin{eqnarray}
\label{gib}
S_{GH,p}= -2(-1)^{p}\epsilon \int_{\partial M_p} \sqrt{-h} H K
d^{D-1}x~.
\end{eqnarray}
Bearing in  mind that
\begin{eqnarray}
\label{p1}
2H \underline{n_{d}}_{;c}g^{a[b}g^{c]d} &=& -(-1)^{p}\epsilon H (K g^{ab}- K^{ab})~, 
\nonumber \\
\label{p2}
4H_{;(c}\underline{n_{d)}}g^{a[b}g^{c]d} &=& -2(-1)^{p}\epsilon [g^{ab}
H_{;c}n^{c} - H^{;(a}n^{b)}]~, \nonumber
\end{eqnarray}
the variation of the full action (supplemented with the boundary term $S_{GH,p}$)
\begin{eqnarray}
\label{acg}
S_{s,p}=\bar{S}_{p}+ S_{GH,p}
\end{eqnarray}
gives
\begin{eqnarray}
\delta S_{s,p} &=& \nonumber \\ \nonumber
&=& - \int_{\partial M_p}d^{D-1}x \sqrt{-h}(-1)^{p} \left\{ \epsilon
\left[(g^{ab}+ \epsilon n^{a}n^{b}) \mathcal{L}_{\vec{n}}H \right. \right.
\\  \nonumber &+& \left. \left. 2 n^{b}h^{ea}H_{,e} + HK h^{ab} -  H K^{ab}
- 2n^{(a} H^{,b)} \right] \right. \\  \nonumber &+& \left. {\chi \over 2} S^{ab} \right\}
\delta g_{ab}\\
&-& 2(-1)^{p} \epsilon  \int_{\partial M_p}d^{D -1}x \sqrt{-h}
K \delta H~,
\end{eqnarray}
where the bulk parts have been omitted. Full variation over the bulk space,
separated by a brane, requires the variation of both of these parts
separately (i.e. first for $p =1$, and then for $p = 2$). This means that the full action
is
\begin{eqnarray}
\label{S12}
\bar{S}_p =S_{s,p=1}+S_{s,p=2} + S_{brane}~,
\end{eqnarray}
where
\begin{equation}
\delta S_{brane} = \int_{\partial M_p} d^{D-1}x \sqrt{-h}{\chi \over 2} S^{ab} \delta
g_{ab}~.
\end{equation}
After varying the total action (\ref{S12}), we obtain the following
junction conditions
\begin{eqnarray}
\label{jc}
&-&(g^{ab}+ \epsilon n^{a}n^{b}) [H_{;c}n^{c}] -
 2 n^{(a}h^{eb)}[H_{,e}] \\ \nonumber &-& [HK] h^{ab} +  [H K^{ab}] + 2n^{(a} [H^{,b)}] =
 \epsilon{\chi \over 2} S^{ab}~,
\end{eqnarray}
where for any quantity $A$ we have defined
\begin{equation}
[A]\equiv A^+ - A^-~.
\end{equation}
After some manipulations, the Eq. (\ref{jc}) can be decomposed into
the following set of conditions:
\begin{eqnarray}
\label{jc2}
[K]&=& 0~, \\
\label{jc21}
S^{ab}n_{a}n_{b}&=& 0~, \\
\label{jc22}
S^{ab}h_{ac}n_{b}&=& 0~, \\
\label{jc23}
-(D-1)[H_{;c}n^{c}]-D[H]K &=& \epsilon {\chi \over 2} S^{ab}h_{ab}~,\\
\label{jc24}
-h_{ab}[H_{;c}n^{c}]-[H]Kh_{ab} &+& [HK_{ab}] \\
&=&
\epsilon {\chi \over 2}S^{cd}h_{ca}h_{db}. \nonumber
\end{eqnarray}
These are the most general junction conditions for $f(R)$ gravity
on the brane. A physical example of a model which allows the discontinuity of the scalaron
$[H] \neq 0$ on the brane is the model which possesses two different cosmological constants
($\Lambda_1$ and $\Lambda_2$) on each side of the brane. In such a case, the Ricci scalar is discontinuous and so the scalaron $H=f'(R)$ is discontinuous, too.

After additional assumption of the continuity
of the scalaron at the brane position, i.e., after taking
\begin{equation}
\label{H=0}
[H]=0~,
\end{equation}
one gets less general junction conditions which were obtained in
Refs. \cite{branef(R),deruelle07}:
\begin{eqnarray}
\label{jc03}
[K]&=&0~, \\
\label{jc3}
[H_{;c}n^{c}]&=&-\epsilon {\chi \over 2(D-1)} S^{ab}h_{ab}~,\\
\label{jc31}
H[K_{ab}]&=&\epsilon {\chi \over 2}\{S^{cd}h_{ca}h_{db}-{h_{ab} h_{cd} \over
(D-1)}S^{cd}\}.
\end{eqnarray}

\section{Gibbons-Hawking boundary term and junction conditions for f(X,Y,Z) gravity.}
\label{sect2}
\setcounter{equation}{0}

In Refs. \cite{clifton,quadratic} a very general fourth-order gravity theory
\bea \label{RRR} S = \frac{1}{\chi}
\int_{M} d^{D}x \sqrt{-g} f(R,R_{ab}R^{ab},R_{abcd}R^{abcd})
\eea
has been studied. It has been shown that such a theory allows basic cosmological solutions such as Einstein, deSitter and G\"odel universes. Motivated by this result, we now develop similar theory, but within the framework of brane universes. As it has already been said, the main problem to start cosmological solutions is to formulate the appropriate Israel junction conditions for such a theory. This is what we are going to do now.

In order to achieve the task, we apply the Gibbons-Hawking boundary term method
to D-dimensional brane universes, given by the action \cite{paper1}
\bea \label{XYZ} S &=& \frac{1}{\chi}
\int_{M} d^{D}x \sqrt{-g} f(R,R_{ab}R^{ab},R_{abcd}R^{abcd}) \nonumber \\
&+& S_{brane} + S_{bulk,p}~
\eea
where $R$, $R_{ab}R^{ab}$, $R_{abcd}R^{abcd}$ are curvature
invariants, and $\chi$ is a constant. This generalized fourth-order theory
includes the first Euler density theory:
$f(R,R_{ab}R^{ab},R_{abcd}R^{abcd})= \chi \kappa_1 R $, as well as the second Euler density
theory (Gauss-Bonnet term): $f(R,R_{ab}R^{ab},R_{abcd}R^{abcd}) = \chi \kappa_2 (R_{abcd}R^{abcd} - 4R_{ab}R^{ab} + R^2)$ ($\kappa_1,\kappa_2=$ const.)
\cite{lovelock}, as special cases. In order to discuss the boundary terms for the action (\ref{XYZ}), we notice that it is still a special case of a more general theory
with the action in the form \cite{kijowski}
\bea \label{r} S_{G} &=& \chi^{-1} \int_{M} d^{D}x \sqrt{-g}
f(g_{ab},R_{abcd}).
\eea
Following the pattern of the $f(R)$ theory given in Section \ref{sect1},
we notice that (\ref{XYZ}) is equivalent to the following second-order theory
\bea \label{equiv}
 S_{I} &=& \chi^{-1} \int_{M} d^{D}x \sqrt{-g} \{ H^{ghij}(R_{ghij}-
 \phi_{ghij}) \nonumber \\
 &+& f(g_{ab},\phi_{cdef}) \}~,
 \eea
where
\bea \label{H} H^{ghij} \equiv {\partial f(g_{ab},\phi_{abcd}) \over
\partial \phi_{ghij}}~, \eea
and
\bea
\label{det2} det \left[{\partial^2 f(g_{ab},\phi_{abcd}) \over
\partial \phi_{ghij} \partial \phi_{klmn}} \right] \neq 0~.
\eea
The condition (\ref{det2}) allows to get an equation of motion of
the $\phi_{klmn}$ field as $\phi_{klmn}=R_{ghij}$ in a similar
manner as for the $Q$ field in the previous Section \ref{sect1}.
In fact, here the tensor $H^{ghij}$ generalizes the scalaron $H=f'(Q)$
of (\ref{equiv2}) onto a tensorial quantity. Because of that, we will call it a tensoron. Varying the equivalent action
(\ref{equiv}), we get the boundary terms in the form
\bea \label{bound1}
\chi^{-1}\{\int_{ \partial M_p} d^{D-1}x \sqrt{-h}  A^{(ab)cd}  n_{d} \delta
g_{ab;c} \\ \nonumber -\int_{ \partial M_p} d^{D-1}x \sqrt{-h}  {A^{(ab)cd}}_{;d} n_{c} \delta
g_{ab}\}~,
\eea
where
\bea
A^{abcd}={1 \over 2} \{H^{acdb} &+& H^{abdc}-H^{cbda} - H^{acbd} \nonumber \\
&-& H^{abcd}+H^{cbad}\}~. \eea
In a special case of the $f(R,R_{ab}R^{ab},R_{abcd}R^{abcd})$ theory, the tensor $A^{abcd}$ reads as
\bea
\label{at}
A^{abcd}&=& f_{X}(g^{ad} g^{cb}-g^{cd} g^{ba}) \nonumber \\
\nonumber &+& f_{Y}(2R^{ad}g^{bc} - R^{cd}g^{ba} -R^{ba}g^{cd})\\
&+& 4f_{Z}R^{acbd}~.
\eea
Finally, an appropriate Gibbons-Hawking boundary term for the
action (\ref{equiv}) is
\bea
\label{gib2}
S_{GH,p} &=& \\
 &-& (-1)^{p} \int_{\partial M_p}
d^{D-1}x \sqrt{-h} A^{(ab)cd}n_{c} n_{d}\mathcal{L}_{\vec{n}}g_{ab}~.\nonumber
\eea
Introducing $X^{c}= A^{(ab)cd}  n_{d} \delta g_{ab}$ into (\ref{di}), and
applying the relations
\be
\label{cond1}
 \mathcal{L}_{\vec{n}}\delta g_{ab} =
 \delta \{\mathcal{L}_{\vec{n}}g_{ab}\} - g_{eb}\{\delta n^{e}\}_{;a}
 - g_{ea}\{\delta n^{e}\}_{;b} 
 \ee
 \be
 \label{cond2}
  A^{(ab)cd} n_{c} n_{d}
 n_{b}=0~,
\ee
where the second equation (\ref{cond2}) is fulfilled for $A^{(ab)cd}$ given by
(\ref{at}), one obtains the following junction conditions for the $f(R,R_{ab}R^{ab},R_{abcd}R^{abcd})$ brane gravity:
\bea
\label{JCXYZ1}
&&[KA^{(ab)cd}] n_{c} n_{d} + [\mathcal{L}_{\vec{n}}A^{(ab)cd}] n_{c} n_{d}
\\ \nonumber &-& \epsilon[A^{(ab)cd}K_{cd}] - g^{ab}[A^{(ef)cd} K_{ef}]n_{c} n_{d}  \\
\nonumber &+& 2 \epsilon [D_{s}A^{(ef)cd}n_{c} n_{d}]h^{s}_{e}h^{(a}_{f}n^{b)}
-  2\epsilon[{A^{(ab)cd}}_{;(c}]n_{d)} = {\chi \over 2} S^{ab}~,
 \\
\label{JCXYZ2}
 \nonumber
 &&n_{b} n_{c}[\mathcal{L}_{\vec{n}}g_{ad}]-
 n_{a} n_{c}[\mathcal{L}_{\vec{n}}g_{db}]-n_{b} n_{d}[\mathcal{L}_{\vec{n}}g_{ac}]\\
 &+&n_{a} n_{d}[\mathcal{L}_{\vec{n}}g_{cb}]=0~.
\eea
These junction conditions coincide with those obtained by the application of a different method in our previous Ref.\cite{paper1} (Eqs. (5.4)-(5.5)),  provided that
\begin{equation}
[A^{(ab)cd}] = 0~,
\end{equation}
i.e., after the assumption that the tensoron is continuous at the
brane position. This is an analogous condition to the condition for the
continuity of the scalaron on the brane (\ref{H=0}) in the $f(R)$ theory of
brane gravity.

\section{Summary}
\label{sum}
\setcounter{equation}{0}

In this paper we derived the most general junction conditions for both $f(R)$ and
$f(R,R_{ab}R^{ab},R_{abcd}R^{abcd})$ braneworld gravities
by the application of the Gibbons-Hawking
boundary term method. We generalized previously obtained junction conditions
for $f(R)$ gravity for the case in which we did not assume the continuity of the
scalaron field $H=f'(Q)$ on the brane. Such a case appears, for example, if one takes two different values of the cosmological constants in the bulk on each side of the brane.
After assuming the continuity of the scalaron, these most general junction conditions
reduce to the ones obtained in earlier references \cite{branef(R),deruelle07}.
Next, we derived the most general junction conditions for the $f(R,R_{ab}R^{ab},R_{abcd}R^{abcd})$ braneworld gravity. Here, we also did not make any
assumption about the continuity of the tensoron field $A^{(ab)cd}$ at the brane position.
Again, we have shown that these junction conditions reduce to those obtained by different methods in our earlier paper \cite{paper1}, provided one assumes the continuity of the tensoron field at the brane. We should stress that even if the
scalaron and tensoron fields are discontinuous on the brane, and
therefore their normal derivatives produce a delta function singularity on the
brane, the quantity which determines a jump of the normal
derivative of the scalaron or the tensoron on the brane is
well-defined (e.g. in the case of delta function having a
singularity at some point $x_0$, a jump of delta function $x_0$
is equal to zero - this comes directly form the definition of the
limit of the function). This obviously means that neither the scalaron nor
the tensoron have to be continuous across the brane, so that the
junction conditions (\ref{jc2})-(\ref{jc24}) and (\ref{JCXYZ1})-(\ref{JCXYZ2}) are
physically realistic.

We think that the method of the Gibbons-Hawking boundary term is more
elegant than other methods of deriving junction conditions, provided one is able to
suggest a correct boundary term. Up to our knowledge, so far, only
the gravity theory of the linear combination of the curvature invariants
$f(R,R_{ab},R_{abcd})=aR^2 + bR_{ab}R^{ab} +cR_{abcd}R^{abcd}$
($a,b,c=$ const.), was studied in the Gibbons-Hawking boundary term
approach \cite{braneR2}.

We consider our result as a basic step in order to write down cosmological equations for $f(R,R_{ab},R_{abcd})$ brane gravity and to check, if basic csomological models are allowed in such a framework. Besides, the discontinuity of the scalaron and tensoron on the brane can be used to describe some other physical cases of the surface layers, such as boundary surfaces separating stars from the surrounding vacuum (singular hypersurfaces of higher order \cite{israel66}).

We also hope that the analysis of the higher-order brane cosmologies
will give some characteristic pattern in statefinder (jerk, kerk/snap,
lerk/crackle, merk/pop) diagnostic of cosmology \cite{statefind}, in a similar manner
as it was given in Ref. \cite{statefR} for $f(R)$ non-brane gravity models. This, however,
will be the matter of a separate paper \cite{future}.

\section{acknowledgments}
We thank Andrei Barvinsky, Salvatore Capozziello, Tomasz Denkiewicz, Krzysztof Meissner
and David Wands for discussions.
We acknowledge partial support of the Polish Ministry of Science
and Higher Education grant No N N202 1912 34 (years 2008-10).


\begin{thebibliography}{99}

\bibitem{wald} Wald R., {\it General Relativity} (University of Chicago Press, 1984).

\bibitem{GH} G.W. Gibbons and S.W. Hawking, Phys. Rev. D {\bf 15}, 2752 (1977).

\bibitem{lutrell} S.W. Hawking and J.C. Lutrell, Nucl. Phys. B{\bf
247}, 250 (1984).

\bibitem{madsen} M. Madsen and J.D. Barrow, Nucl. Phys. B{\bf
323}, 242 (1989).

\bibitem{barvinsky} A.D. Barvinsky and S.N. Solodukhin, Nucl.
Phys. B {\bf 479}, 305 (1996).

\bibitem{lovelock} D. Lovelock, J. Math. Phys. {\bf 12}, 498 (1971).

\bibitem{briggs} C.C. Briggs, gr-qc/9808050.

\bibitem{bunch81} T.S. Bunch, Journ. Phys. A {\bf 14}, L139 (1981).

\bibitem{surface} F. M\"uller-Hoissen, Phys. Lett. B {\bf 163}, 106
(1985); R.C. Myers, Phys. Rev. D {\bf 36}, 392 (1987).

\bibitem{davis} S.C. Davis, Phys. Rev. D {\bf 67}, 024030 (2003).

\bibitem{gravanis} E. Gravanis and S. Willinson, Journ. Math. Phys. {\bf 47}, 2503 (2006);
Phys. Rev. D{\bf 75}, 084025 (2007).

\bibitem{olea} O. Mi\v{s}kovi\'{c} and R. Olea, JHEP {\bf 0710},
028 (2007).

\bibitem{hw} P. Ho\v{r}ava and E. Witten, Nucl. Phys. B{\bf 460}
(1996), 506; {\it ibid} B{\bf 475}, 94.

\bibitem{RS} L. Randall and R. Sundrum, Phys. Rev. Lett., {\bf 83}, 3370
(1999); L. Randall and R. Sundrum, {\it ibidem}, {\bf 83}, 4690
(1999).

\bibitem{brane}
M. Visser, Phys. Lett. B{\bf 159}, 22 (1985);
N. Arkani-Hamed, S. Dimopoulos, and G. Dvali,
Phys. Lett. B{\bf 516}, 70 (1998); I. Antoniadis, N. Arkani-Hamed, S.
Dimopoulos, G. Dvali, Phys. Lett. B{\bf 436}, 257 (1998);
N. Arkani-Hamed, S. Dimopoulos, and G. Dvali,
Phys. Rev. D{\bf 59}, 086004 (1999).

\bibitem{visser} M. Visser, {\it Lorentzian Wormholes}
(Springer-Verlag, 1996).

\bibitem{israel66} W. Israel, Nuovo Cimento B {\bf 44}, 1 (1966).

\bibitem{GaussCod} P. Bin\'etruy, C. Deffayet and D. Langlois, Nucl.
Phys. B{\bf 565}, 269 (2000);
P. Bin\'etruy, C. Deffayet and D. Langlois, Phys.
Lett. B{\bf 477}, 285 (2000);
M. Sasaki, T. Shiromizu and K. Maeda, Phys.
Rev. D{\bf 62}, 024008 (2000);
T. Shiromizu, K.I. Maeda, and M. Sasaki, Phys. Rev. D{\bf 62}, 024012
(2000);
S. Mukhoyama, T. Shiromizu and K. Maeda,
Phys. Rev. D{\bf 62}, 024028 (2000).

\bibitem{abdalla} M. Arik and D. \c{C}iftci, Gen. rel. Grav. {\bf 37}, 2211 (2005); M.C.B. Abdalla, M.E.X. Guimar$\tilde{a}$es, and
J.M. Hoff de Silva, hep-th/0711.1254.

\bibitem{deruelle00} N. Deruelle and T. Dole\v{z}el, Phys. Rev.
D{\bf 62}, 103502 (2000).

\bibitem{charmousis} C. Charmousis, J.F. Dufaux, Class. Quantum Grav. {\bf 19}, 4671 (2002).


\bibitem{jim} J.F. Dufaux, J.E. Lidsey, R. Maartens, and M. Sami, Phys. Rev. D {\bf 70}, 083525 (2004).

\bibitem{lidsey} J.E. Lidsey, Ann. Phys. (Leipzig) {\bf 15}, 277
(2006).

\bibitem{maeda} H. Maeda, V. Sahni, Yu. Shtanov, Phys. Rev. D {\bf 76}, 104028 (2007).

\bibitem{meissner01} K.A. Meissner and M. Olechowski, Phys. Rev.
Lett. {\bf 86}, 3708 (2001).

\bibitem{paper1} A. Balcerzak and M.P. D\c{a}browski, Phys. Rev.
D{\bf 77}, 023524 (2008).

\bibitem{f(R)} A.A. Starobinsky, Phys. Lett. B {\bf 91}, 99 (1980);
G. Magnano and L.M. Soko{\l }owski, Phys. Rev. D {\bf 50}, 5039 (1994);
T.P. Sotiriou, Class. Quantum Grav. {\bf 23}, 5117 (2006);
T. Chiba, T. Kobayashi, M. Yamaguchi, and J. Yokoyama, Phys. Rev. D {\bf 75}, 043516 (2007);
G.J. Olmo, Phys. Rev. Lett. {\bf 98}, 061101 (2007); G.J. Olmo, Phys. Rev. D {\bf 75},
023511 (2007);
S. Capozziello, V.F. Cardone, and A. Troisi, Phys. Rev. D {\bf 71}, 043503 (2005);
S. Capozziello, S. Nojiri, S.D. Odintsov, and A. Troisi, Phys. Lett.
B {\bf 639}, 135 (2006);
S. Capozziello and R. Garatini, Class. Quantum Grav. {\bf 24}, 1627 (2007);
L. Amendola, D. Polarski, and S. Tsujikawa, Phys. Rev. Lett. {\bf 98}, 131302
(2007); S. Nojiri and S.D. Odintsov, arXiv: 0804.3519; 0807.0685;
T. Faulkner, M. Tegmark, E.F. Bunn, and Y. Mao, Phys. Rev. D {\bf 76},
063505 (2007); N. Lanahan-Tremblay and V. Faraoni,
hep-th/0709.4414; Q. Exirifard, gr-qc/0708.0662;
S. Capozziello, V.F. Cardone, and A. Troisi,
astro-ph/0604435.

\bibitem{RMP} J. Sotiriou and V. Faraoni, arXiv:0805.1726.

\bibitem{clifton} T. Clifton and J.D. Barrow, Phys. Rev. D {\bf 72}, 123003 (2005);
T. Clifton and J.D. Barrow, Class. Quantum Grav. {\bf 23}, 2951 (2006).

\bibitem{quadratic} J.D. Barrow and S. Hervik, Phys. Rev. D {\bf 73}, 023007 (2006);
{\it ibidem}, {\bf 74}, 124017 (2006).

\bibitem{afrolov} A. Frolov, astro-ph/0803.2500.

\bibitem{borzeszkowski} H.H. v. Borzeszkowski and V.P. Frolov,
Ann. Phys. (Leipzig) {\bf 7}, 285 (1980).

\bibitem{branef(R)} M. Parry, S. Pichler, and D. Deeg,
JCAP {\bf 0504}, 014 (2005).

\bibitem{deruelle07} N. Deruelle, M. Sasaki, and Y. Sendouda,
gr-qc/0711.1150.

\bibitem{braneR2} S. Nojiri and S.D. Odintsov, JHEP {\bf 0007},
049 (2000); S. Nojiri, S.D. Odintsov and S. Ogushi, Phys. Rev. D
{\bf 65}, 023521 (2001).

\bibitem{kijowski} A. Jakubiec and J. Kijowski, Phys. Rev. D {\bf
37}, 1406 (1988).

\bibitem{statefind} V. Sahni, T.D. Saini, A.A. Starobinsky, and U. Alam, JETP Lett.
               {\bf 77}, 201 (2003); M. Visser, Class. Quantum Grav. {\bf 21}, 2603 (2004);
               R.R. Caldwell and M. Kamionkowski, JCAP 0409, 009 (2004);
               M.P. D\c{a}browski, Phys. Lett. B{\bf 625}, 184 (2005);
               M.P. D\c{a}browski and T. Stachowiak, Annals of Physics
               (New York) {\bf 321}, 771 (2006); M. Dunajski and
               G.W. Gibbons, arXiv: 0807.0207.

\bibitem{statefR} N.J. Poplawski, Phys. Lett. B{\bf 640}, 135 (2006);
Class. Quantum. Grav. {\bf 24}, 3013 (2007); S. Capozziello, V.F. Cardone, and V.
Salzano, astro-ph/0802.1583.

\bibitem{future} A. Balcerzak and M.P. D\c{a}browski - in preparation.















\end{thebibliography}
\end{document}